\documentclass[preprint,english,showpacs,11pt,floatfix]{revtex4}

\usepackage{babel,amsmath,amssymb,dcolumn}
\usepackage[dvips]{graphics}
\usepackage{hyperref}

\usepackage{amsfonts}
\usepackage{amssymb}
\usepackage[dvips]{graphicx}
\usepackage{latexsym}

\usepackage{dcolumn}
\usepackage{bm}

\newcommand{\be}{\begin{equation}}
\newcommand{\ee}{\end{equation}}
\newcommand{\no}{\nonumber}

\begin{document}

\title{Transition of the dark energy equation of state in an interacting
holographic dark energy model}

\author{Bin Wang}
\email{wangb@fudan.edu.cn} \affiliation{Department of Physics,
Fudan University, Shanghai 200433, People's Republic of China }

\author{Yungui Gong} \email{gongyg@cqupt.edu.cn}
\affiliation{College of Electronic Engineering, Chongqing
University of Posts and Telecommunications, Chongqing 400065,
China}

\author{Elcio Abdalla}
\email{eabdalla@fma.if.usp.br} \affiliation{Instituto de Fisica,
Universidade de Sao Paulo, C.P.66.318, CEP 05315-970, Sao Paulo,
Brazil}

\begin{abstract}
A model of holographic dark energy with an interaction with matter fields
has been investigated. Choosing the future event horizon as an IR cutoff,
we have shown that the ratio of energy densities can vary with time. With
the interaction between the two different constituents of the universe, we
observed the evolution of the universe, from early deceleration to late
time acceleration. In addition, we have found that such an interacting
dark energy model can accommodate a transition of the dark energy from a
normal state where $w_D>-1$ to $w_D<-1$ phantom regimes. Implications of
interacting dark energy model for the observation of dark energy
transition has been discussed.
\end{abstract}


\maketitle

The present acceleration of the universe expansion has been well
established through numerous and complementary cosmological
observations \cite{1}. A consistent picture has indicated that
nearly three quarters of our universe consists of a ``dark
energy", which is responsible for the accelerated expansion.
However, the nature of such a dark energy is still rather
uncertain. Explanations have been sought within a wide range of
physical phenomena, including a cosmological constant, exotic
fields, a new form of the gravitational equation, new geometric
structures of spacetime, etc \cite{2,asaa}. Recently, a new model
stimulated by the holographic principle has been put forward to
explain the dark energy \cite{3}. According to the holographic
principle, the number of degrees of freedom of a physical system
scales with the area of its boundary. In this context, Cohen et al
\cite{cohen} suggested that in quantum field theory a short
distance cutoff is related to a long distance cutoff due to the
limit set by formation of a black hole, which results in an upper
bound on the zero-point energy density. In line with this
suggestion, Hsu and Li \cite{Hsu}\cite{3} argued that this energy
density could be viewed as the holographic dark energy density
satisfying $\rho_D= 3c^2/L^2$, where $c^2$ is a constant, $L$ is
an IR cutoff in units $M_p^2=1$. Li \cite{3} discussed three
choices for the length scale $L$ which is supposed to provide the
IR cutoff, such as the Hubble radius, the particle horizon and the
event horizon. He demonstrated that only identifying $L$ with the
radius of the future event horizon, we can get the correct
equation of state of dark energy and obtain the desired
accelerating universe. Besides other applications of the
holographic principle in cosmology \cite{4}, the holographic dark
energy model is a new example showing that holography is an
effective way to investigate cosmology. Related works on the
holographic dark energy can be found in \cite{5}. The holographic
dark energy model is found in consistent with the observational
data \cite{observation}.

Available models of dark energy differ in the equation of state
parameter $w_D$ as well as the variation of $w_D$ itself during
the evolution of the universe. The cosmological constant with
$w_D=-1$, is located at a central position among dark energy
models both in theoretical investigation and in data analysis
\cite{6}. In quintessence, K-essence, Chaplygin gas and
holographic dark energy models  \cite{2,3}, $w_D$ stays bigger
than $-1$, and cannot cross $-1$. The phantom models of dark
energy have $w_D<-1$  \cite{7,asaa}. Recently, the analysis of the
type Ia supernova data indicates that the time varying dark energy
gives a better fit than a cosmological constant \cite{8}. These
analysis mildly favor the evolution of the dark energy parameter
$w_D$ from $w_D>-1$ to $w_D<-1$ at recent stage. Although the
galaxy cluster gas mass fraction data do not support the
time-varying $w_D$ \cite{chen}, theoretical attempts toward the
understanding of the $w_D$ corssing $-1$ phenomenon have been
started. Some dark energy models, such as the one containing a
negative kinetic scalar field and a normal scalar field \cite{9},
or a single scalar field model \cite{10}, a Gauss-Bonnet brane
world with induced gravity \cite{11} and the generalized
implicitly defined dark energy equation of state model \cite{12},
have been constructed to gain insight into the occurrence of the
transition of the dark energy equation of state and the mechanism
behind this transition. Other studies on the $w_D$ crossing $-1$
can be found in \cite{other}.

Most discussions on dark energy rely on the fact that its
evolution is independent of other matter fields. Given the unknown
nature of both dark energy (DE) and dark matter (DM), which are
two major contents of the universe, one might argue that an
entirely independent behavior of DE is very special \cite{13}.
Studies about the interaction between DE and DM have been carried
out in \cite{13,14,15,16}. It was argued that the interaction will
influence the perturbation dynamics and could be observable
through the lowest multi-poles of the CMB spectrum \cite{14}.
Recently, by considering the interaction between DE and DM in the
holographic DE model, it was even found that the Hubble scale
might be used as the IR cutoff to explain the
acceleration of our universe \cite{16}. In
this work, we are going to
extend the inclusion of interaction between DE and DM into the
holographic DE model with the future event horizon as the IR
cutoff. The ratio of energy densities can be varied with time. As
a result, we find that with the interaction between DE and DM, the
model can give an early deceleration and a late time acceleration.
In addition, in this holographic model, the appropriate coupling
between DE and DM accommodates the transition of the DE equation
of state from $w_D>-1$ to $w_D<-1$. This property could serve as
an observable feature of the interaction between DE and DM in
addition to its influence on the small $l$ CMB spectrum argued in
\cite{14}.

The total energy density is $\rho=\rho_m+\rho_D$, where $\rho_m$
is the energy density of matter and $\rho_D=3c^2 R_E^{-2}$ is the
dark energy density. Here, the Planck mass is taken as unit. We
followed \cite{3} by choosing the future event horizon,
$R_E=a\int_a^{\infty}\frac{dx}{Hx^2}$, as the IR cutoff, where
$c^2$ is a constant. The total energy density satisfies a
conservation law. However since we consider the interaction
between DE and DM, $\rho_m$ and $ \rho_D$ do not satisfy
independent conservation laws, they instead satisfy  \be
\dot{\rho}_m+3H\rho_m=Q \label{1} \ee and \be
\dot{\rho}_D+3H(1+w_D)\rho_D=-Q\quad ,\label{2} \ee where $w_D$ is
the equation of state of DE, $Q$ denotes the interaction term and
can be taken as $Q=3b^2 H\rho$ with $b^2$ the coupling constant
\cite{16}. This expression for the interaction term was first
introduced in the study of the suitable coupling between a
quintessence scalar field and a pressureless cold dark matter
field \cite{13}. The choice of the interaction between both
components was to get a scaling solution to the coincidence
problem such that the universe approaches a stationary stage in
which the ratio of dark energy and dark matter becomes a constant.
In the context of holographic DE model, this form of interaction
was derived from the choice of Hubble scale as the IR cutoff \cite{16}.

Taking the ratio of energy densities as $r=\rho_m/\rho_D$, from
(\ref{1}) and (\ref{2}) we have
\be \dot{r}=3b^2
H(1+r)^2+3Hrw_D\quad.\label{3} \ee

In \cite{16}, the IR cutoff was chosen as being the Hubble scale, which
leads to a constant $r$. We choose the
future event horizon as the IR cutoff so that $r$ is
no longer a constant in the holographic DE model.

Using the Friedmann equation, $\Omega_m+\Omega_D=1$, where
$\Omega_m=\rho_m/(3H^2)$ and $\Omega_D=\rho_D/(3H^2)$, we have
$r=(1-\Omega_D)/\Omega_D$ and $\dot{r}=-\dot{\Omega}_D/\Omega_D^2$.
Combining with (\ref{3}), we get the equation of state of DE,
\begin{eqnarray}
w_D & = & [-\dot{\Omega}_D/\Omega_D^2-3b^2H(1+r)^2]/(3Hr) \no \\ & = &
-\frac{\Omega_D'}{3\Omega_D(1-\Omega_D)}-\frac{b^2}{\Omega_D(1-\Omega_D)}
\quad , \label{4}
\end{eqnarray}
where the dot is the derivative with respect to time and the prime is the
derivative with respect to $x=\ln a$.

From the Friedmann equation, the future event horizon can be expressed as
$R_E=c\sqrt{1+r}/H=a\int_t^{\infty} dt/a$. Taking the derivative with
respect to $t$ and using the definitions of $r$ and $ \dot{r}$ as well as
equation (\ref{4}), we arrive at
\be
\frac{\Omega_D'}{\Omega_D^2}=(1-\Omega_D)[\frac{1}{\Omega_D}+
\frac{2}{c\sqrt{\Omega_D}}-\frac{3b^2}{\Omega_D(1-\Omega_D)}]\quad . \label{5}
\ee
Back to (\ref{4}) we get now
\be
w_D =  -1/3-2\sqrt{\Omega_D}/(3c)-b^2/\Omega_D\quad .
\ee
Neglecting the interaction between DE and DM, (\ref{5}) is equivalent to
equation (18) of Li's paper in \cite{3}. We will show that the interaction between DE and DM
brings rich physics.

The deceleration parameter can be expressed in this model as
\begin{eqnarray}
q& = & -\ddot{a}a/a^2=-\dot{H}/H^2-1 \no \\
& = & 1/2-3b^2/2-\Omega_D/2-\Omega_D^{3/2}/c\quad , \label{6}
\end{eqnarray}
where $\dot{H}/H^2=3(-1-w_D-r)\Omega_D/2$.

With eqs (\ref{4}-\ref{6}), we are in a position to investigate the
evolution of the DE and its influence on the expansion of the universe.

Since the DE plays a more  important role in the evolution of the universe
with the flow of cosmological time, we require $\Omega_D'>0$, which leads to
\be
b^2<b^2_{max}=(1-\Omega_D)(1+2\sqrt{\Omega_D}/c)/3\quad .\label{7}
\ee

\begin{figure}[!hbtp]
\begin{center}
\includegraphics[width=10cm]{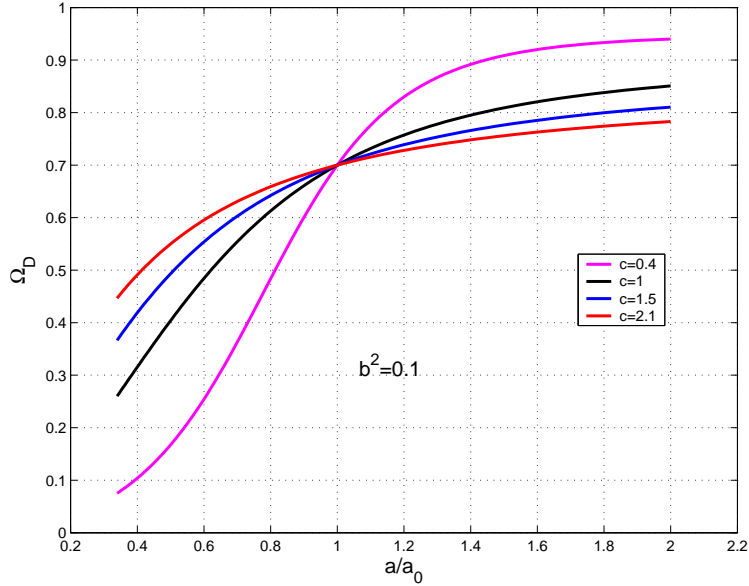}
\end{center}
\caption{Evolution of the DE for a fixed interaction parameter with DM
  ($b^2$) but for different values of the constant $c$.   }
\end{figure}
\begin{figure}[!hbtp]
\begin{center}
\includegraphics[width=10cm]{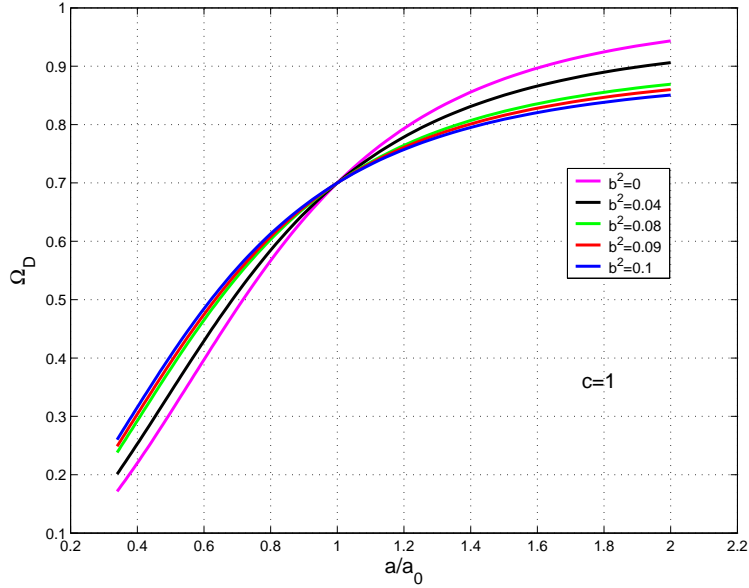}
\end{center}
\caption{Evolution of the DE with a fixed constant $c$ but different
  values for the coupling with DM. }
\end{figure}

The behavior of the DE evolution puts an upper limit on the interaction
between DE and DM, which has not been observed in the study of the
holographic DE model by taking the Hubble scale as IR cutoff \cite{16}.
$b^2_{max}$ decreases with the increase of $c$. Choosing two different
values of $c$ with $c_1<c_2$, their corresponding largest allowed values
$b^2_{max}$ are $b^2_{1-max}$ and $b^2_{2-max}$ and satisfy $b^2_{1-max}>
b^2_{2-max}$. Taking $b^2_s$ as a common allowed value of the coupling
for $c_1$ and $c_2$ cases, it is easy to see that $b^2_s$ is closer to
$b^2_{2-max}$ than  $b^2_{1-max}$, which means that in the large $c$ case
the same allowed coupling $b^2$ is stronger than in the small $c$ case.

The dependences of the evolution of DE with respect to the constants $b^2,
c$ are shown in Fig. 1 and Fig. 2, respectively. From Fig. 1 we learn that for
the fixed coupling between DE and DM within the allowed range (\ref{7}),
the DE starts to be effective earlier when $c$ is larger. Since the
same $b^2$ corresponds to a stronger coupling between DE and DM for bigger
$c$ cases, $\Omega_D$ will tend to a smaller value at late
stage when $c$ is bigger due to this stronger coupling.

For a fixed $c$, the dependence of $\Omega_D$  on $b^2$ is shown in Fig. 2.
When $b^2$ is larger, at early stage, the DE starts to be effective
earlier. However at a later stage, since the coupling between DE and DM
becomes stronger, $\Omega_D$ approaches a smaller value for a larger $b^2$.

Including the interaction, our model naturally shows that our universe has
an accelerated expansion in the late stage and on the other hand it also
displays a deceleration in the early era. In Fig. 3 we show that for the
same coupling between DE and DM, the acceleration starts earlier for
larger $c$, since the DE develops earlier for larger $c$ with the same
coupling $b^2$. In Fig. 4 we show that for the same $c$, the acceleration
starts earlier for larger $b^2$. This is also due to the fact that for the
same $c$, DE develops earlier for bigger $b^2$ as shown in Fig. 2.

\begin{figure}[!hbtp]
\begin{center}
\includegraphics[width=10cm]{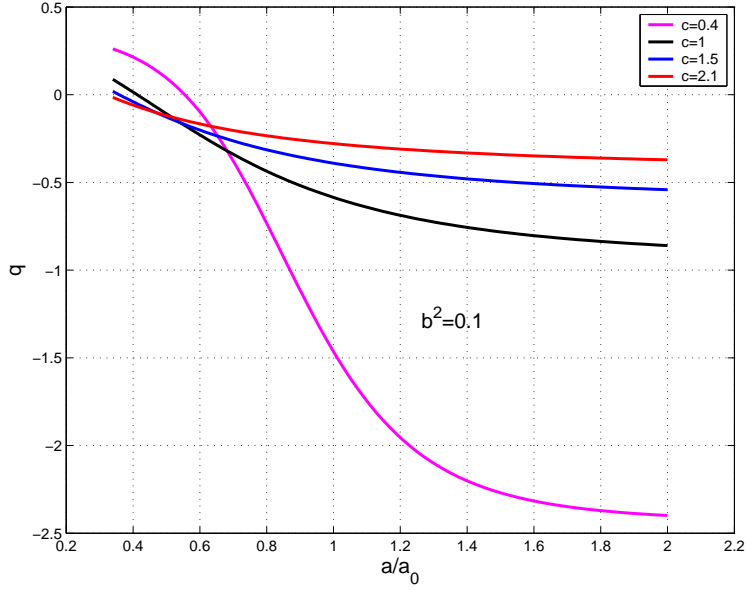}
\end{center}
\caption{ Dependence of the deceleration parameter on the constant $c$ for
  a fixed coupling between DE and DM. }
\end{figure}

\begin{figure}[!hbtp]
\begin{center}
\includegraphics[width=10cm]{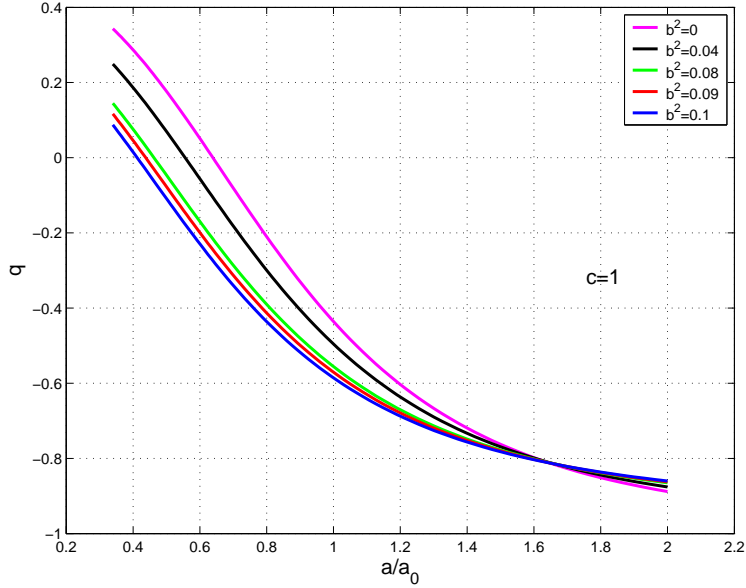}
\end{center}
\caption{ Dependence of the deceleration parameter on the coupling between
  DE and DM for a fixed constant $c$. }
\end{figure}

We now discuss the equation of state of the DE with the interaction
between DE and DM. From (\ref{6}) we learnt that $w_D$ has a maximum value
$w_{D-max}=-1/3-(b/\sqrt{3}c)^{2/3}$ at $\Omega_{D-cr}=(3b^2c)^{2/3}$. To
allow the DE transition as indicated by recent observations with $w_D$
crossing the border $-1$ and $w_D<-1$ at present stage, we need
$b^2>b^2_{cr}=2\Omega_D(1-\sqrt{\Omega_D}/c)/3$, where $b^2_{cr}$ is got
from $w_D=-1$.
Such $b^2_{cr}$ should be in the reasonable range
(\ref{7}). Thus we get $c<2\sqrt{\Omega_D}/(3\Omega_D-1)$. Meanwhile
$b^2_{cr}$ should be positive, which leads to $c>\sqrt{\Omega_D}$.
According to the holographic principle, the entropy of the universe is bounded by $S=\pi R_E^2$, where
$R_E=c/(H\sqrt{\Omega_D})$. If we require the entropy of the universe do not decrease, we need
$\dot{R}_E=c/\sqrt{\Omega_D}-1\geq 0$, thus we get $c\geq \sqrt{\Omega_D}$. Therefore the
lower bound on $c$ is exactly the requirement of the second law of
thermodynamics discussed in \cite{3}. Since the DE evolves independently
of DM in cases ($b^2=0$), $w_D<-1$ requires $c<\sqrt{\Omega_D}$, which
violates the second law. Thus in the holographic model with entirely
indepedent behavior between DE and DM, it is impossible to have  $w_D$
crossing $-1$ allowing the phantom energy to exist in the late stage of
the universe.

Further examining the function $b^2_{cr}$, we see that it has a maximum
value $b^2_{cr-max}=8c^2/81$ at $\Omega_D=4c^2/9$. In order to have
$w_D<-1$, we need $b^2>b^2_{cr}$. However the coupling $b^2$ cannot be
arbitarily large, since when $b^2>b^2_{cr-max}$ and $w_D<-1$ always. The
upper bound on $b^2$ can also be got by requiring $w_{D-max}>-1$ thus
allowing $w_D$ to cross $-1$ and to stay below $-1$ later.
In order to have $w_D<-1$ with $\Omega_{D0}=0.7$, such a maximum value
$b^2_{cr-max}$ should appear early when $\Omega_D=4c^2/9<0.7$. Combining
constraints on $c$, we have
\be
\sqrt{\Omega_D}<c<1.255\quad .
\ee

Now let us focus on the parameter space of $b^2$, which we have already
obtained, $b^2_{cr}<b^2<8c^2/81$. Meanwhile taking account of the
behavior of the $w_D$ function, we also require
$\Omega_{D-cr}<\Omega_{D0}$, which leads to $b^2<\Omega_{D0}^{3/2}/(3c)$,
allowing the $w_D$ transition from $w_D>-1$ to $w_D<-1$ in recent times.
Using $\Omega_{D0}=0.7$, the combined constraint on $b^2$ reads
\be
1.4(1-\sqrt{0.7}/c)/3<b^2<8c^2/81\quad ,
\ee
to accommodate $w_D$ crossing $-1$.

For a fixed $c$, the dependence of $w_D$ on $b^2$ is shown in Fig. 5. We
see that for larger $b^2$, $w_D$ crosses $-1$ earlier. For a too
small $b^2$, $w_D$ crosses $-1$ too late. Actually these small $b^2$ are
over the lower bound of (10) and are not consistent with the
observation.

\begin{figure}[!hbtp]
\begin{center}
\includegraphics[width=10cm]{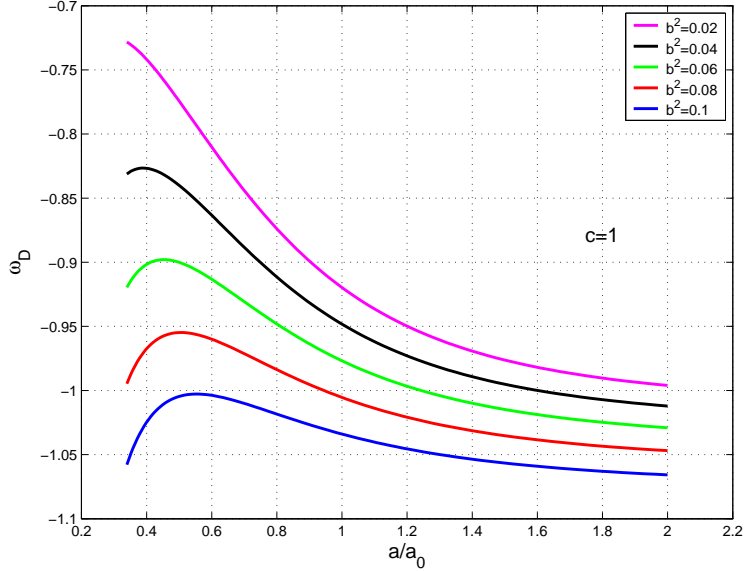}
\end{center}
\caption{ Behavior of the equation of state of DE for different couplings
  between DE and DM. }
\end{figure}

From the future precise observation of the location of the transition of
the $w_D$ from $w_D>-1$ to $w_D<-1$, it is possible to understand the
interaction between DE and DM. In the future this could serve as another
observable feature of the interaction between DE and DM in addition to the
low CMB spectrum discussed in \cite{14}.

We have also fitted our model with the golden SN data. We got
$\Omega_{m0}=0.39^{+0.14}_{-0.16}$, $b^2=0.00^{+0.11}_{-0.00}$,
$c=0.40^{+0.75}_{-0.30}$ and $\chi^2=174.44$. If we fix $c=1$, we have $\Omega_{m0}=0.26^{+0.14}_{-0.05}$, $b^2=0.00^{+0.12}_{-0.00}$,
and $\chi^2=177.08$. This shows that our model is consistent with the SN
data and the constraint on parameter spaces we discussed are compatible
with the observations.

In summary, we have studied the holographic dark energy model with an
interaction between DE and DM. Choosing the future event horizon as the IR
cutoff, we obtained a generalized model with the time dependent ratio of
energy densities which cannot be realized just by adopting the Hubble
scale as the IR cutoff. With the interaction between DE and DM, a richer
physics arises. In addition to showing the comprehensive history of the
evolution of our universe from the early deceleration to late
acceleration, we have also found that the interaction between DE and DM
can accommodate the transition of the equation of state of DE from
$w_D>-1$ to $w_D<-1$. We have constrained the parameter space of our model
to explain the observations. We argued that in addition to some
observational feature for small $l$ CMB spectrum \cite{14}, the
interaction between DE and DM could also be observed by the $w_D$ crossing
$-1$ behavior in the future. Comparison to the golden SN data has been
made, and we showed that our model is also consistent with such observations.

\begin{acknowledgments}
This work was partially supported by  NNSF of China, Ministry of
Education of China, Ministry of Science and Technology of China
under grant No. NKBRSFG19990754 and Shanghai Education Commission.
Y. Gong's work was supported by NNSFC under grant No. 10447008,
CSTC under grant No. 2004BB8601, CQUPT under grant No. A2004-05
and SRF for ROCS, State Education Ministry. E. Abdalla's work was
partially supported by FAPESP and CNPQ, Brazil.  B. Wang would
like to acknowledge the associate programme in ICTP where the work
was done. \end{acknowledgments}



\begin{thebibliography}{99}

\bibitem{1} A.G. Riess, et al., Astron. J.  116 (1998) 1009;
  S. Perlmutter, et al.,   Astrophys. J.  517 (1999) 565;
  S. Perlmutter, et al.,   Astrophys. J.  598
  (2003) 102;  P. de Bernardis, et al.,   Nature  404 (2000) 955.

\bibitem{2} T. Padmanabhan, Phys. Rep.  380 235 (2003); P. J.
E.Peebles,  B. Ratra,  Rev. Mod. Phys. 75 559 (2003); V. Sahni,
astro-ph/0403324 and
  references therein.

\bibitem{asaa} A. Melchiorri, L. Mersini-Houghton, C.J. Odman, M.
Trodden, Phys. Rev. D 68 (2003) 043509; E. Gunzig and A. Saa, Int.
J. Mod. Phys.
 D 13 (2004) 2255; F.C. Carvalho and A. Saa, Phys. Rev. D 70 (2004) 087302.

\bibitem{3}  M. Li,   Phys. Lett.  B 603 (2004) 1; Q.G. Huang and
M. Li,  JCAP  0408 (2004) 013.

\bibitem{cohen} A. Cohen, D. Kaplan and A. Nelson,   Phys. Rev.
Lett.  82 (1999) 4971.

\bibitem{Hsu} S. D. H. Hsu,  Phys. Lett.   B 594 (2004) 13.

\bibitem{4}  B. Wang, E. Abdalla and T. Osada,   Phys. Rev. Lett.
 85 (2000) 5507;  P. Horava and D. Minic,   Phys. Rev.
Lett.  85 1610 (2000); T. Banks and W. Fischler astro-ph/0307459;
B. Wang and E. Abdalla, Phys. Rev. D 69 (2004) 104014; R.G. Cai,
JACP 0402 (2004) 007; D.A. Lowe and D. Marolf, Phys. Rev. D 70
(2004) 026001.

\bibitem{5}
B. Wang, E. Abdalla and R. K. Su,
 Phys. Lett.  B 611 (2005) 21; Y. S. Myung, hep-th/0501023;
Y.G. Gong and Y.Z. Zhang, hep-th/0505175; J.Y. Shen, B. Wang, E.
Abdalla and R.K. Su,   Phys. Lett.  B 609 (2005) 200; Z.Y. Huang,
B. Wang, E. Abdalla and R.K. Su, hep-th/0501059; E. Elizalde, S.
Nojiri, S.D. Odintsov and P. Wang, hep-th/0502082.

\bibitem{observation} Q.G. Huang
  and Y.G. Gong,    JCAP  0408 (2004) 006; Y.G. Gong, B. Wang and Y. Z. Zhang, hep-th/0412218;
X. Zhang, astro-ph/0504586.

\bibitem{6} S. Weinberg, Rev. Mod. Phys.  61 (1989) 1; N.
Straumann, astro-ph/0203330; T. Padmanabhan, hep-th/0406060.

\bibitem{7}  R.R. Caldwell, Phys. Lett.   B 545 (2002) 23.

\bibitem{8}  U. Alam, V. Sahni and A. A. Starobinsky, JCAP
    0406 (2004) 008; D. Huterer and A. Cooray, Phys. Rev. D 71
  (2005) 023506; Y.G. Gong, Int. J. Mod. Phys. D 14 (2005) 599;
  Y.G. Gong, Class. Quantum Grav. 22 (2005) 2121; Y. Wang and M. Tegmark,
  astro-ph/0501351; Y.G. Gong, astro-ph/0502262.

\bibitem{chen} G. Chen and B. Ratra, Astrophys. J. 612 (2004) L1.

\bibitem{9} B. Feng, X. L.Wang and X. M. Zhang, Phys. Lett.
 B 607 (2005) 35; W. Hu, Phys. Rev. D 71 (2005) 047301; Z.K. Guo, Y.S. Piao, X.M. Zhang and
  Y.Z. Zhang, Phys. Lett.  B 608 (2005) 177;
X.F. Zhang, H.Li, Y.S. Piao and X. Zhang, astro-ph/0501652.

\bibitem{10} M.Z. Li, B. Feng and X.M. Zhang, hep-ph/0503268.

\bibitem{11} R.G. Cai, H.S. Zhang and A.Z. Wang, hep-th/0505186.

\bibitem{12} H. Stefancic, astro-ph/0504518.

\bibitem{other} P. Singh, gr-qc/0502086; S. Nojiri and S.D.
Odintsov,  Phys. Lett.  B
    562 (2003) 147; S. Nojiri and S.D. Odintsov, Phys. Rev. D 70 (2004) 103522;
A. Vikman, Phys. Rev. D 71 (2005) 023515; A. Anisimov, E. Babichev
and  A. Vikman, JCAP 0506 (2005) 006; S. Nojiri, S.D. Odintsov and
S. Tsujikawa, Phys. Rev. D 71 (2005) 063004.

\bibitem{13} L. Amendola, Phys. Rev. D 62 (2000) 043511;
  W. Zimdahl, D. Pav'on and L.P. Chimento, Phys. Lett.  B
    521 (2001) 133; W. Zimdahl and D. Pavon, Gen. Rel. Grav. 35 (2003) 413;
    L.P. Chimento, A.S. Jakubi, D. Pavon and W. Zimdahl,
    Phys. Rev. D 67 (2003) 083513.

\bibitem{14} W. Zimdanl, gr-qc/0505056.

\bibitem{15} G. Mangano, G. Miele and
  V. Pettorino, Modern. Phys. Lett.  A 18 (2003) 831; B. Guberina, R. Horvat
and H. Stefancic, JCAP 0505 (2005) 001;
  G. Farrar and P.J.E, Peebles, Astrophys. J.  604 (2004) 1;
    S. del Campo, R. Herrera, and D. Pavon,  Phys. Rev.  D 70 (2004)
      043540; R.G. Cai and  A.Z. Wang, JCAP 0503 (2005) 002; M. Doran and J. Jackel,
  Phys. Rev. D 66 (2002) 043519; R. Horvat, Phys. Rev. D
  70 (2004) 087301.

\bibitem{16}  D. Pavon and W. Zimdahl, gr-qc/0505020.





\end{thebibliography}
\end{document}